\documentstyle{article}
\begin{document}
\title{
Renormalisability of the SU(n) Gauge Theory \\ with Massive Gauge Bosons}
\author{Ze-sen Yang, Zhining Zhou, Yushu Zhong and Xianhui Li\\
Department of Physics,  Peking University, Beijing 100871, CHINA }
\date{\today}
\maketitle
\begin{abstract}
   The problem of renormalisability of the SU(n) theory with massive gauge
bosons is reinverstigated in the present work. We expound that the
quantization under the Lorentz condition caused by the mass term of the
gauge fields leads to a ghost action which is the same as that of the usual
SU(n) Yang--Mills theory in the Landau gauge. Furthermore, we clarify that
the mass term of the gauge fields cause no additional complexity to the
Slavnov-Taylor identity of the generating functional for the regular vertex
functions and does not change the equations satisfied by the divergent part
of this generating functional. Finally, we prove that the renormalisability
of the theory can be deduced from the renormalisability of the Yang--Mills
theory.
\end{abstract}
\newpage
\begin{center}
{\bf I}.\ \ Introduction
\end{center} \par \ \par
    As is well known, a negative answer to the question of whether a SU(n)
theory with massive gauge bosons is renormalisable was commonly accepted
even before knowing the Faddeev-Popov-De Witt method [1--3] to quantize
the usual SU(n) Yang-Mills gauge theory. However, for various reasons
including the lack of experimental evidence for the Higgs bosons of the
 SU$_L$(2)$\times$U$_Y$(1) electroweak theory, this issue was repeatedly
studied (see for example Refs. [4--10]) and several approaches have been
developed for finding a positive answer. The authors of Refs. [4--6],
studied intersting models where some terms in the action were introduced
as an assumption. In Ref. [7] the mass term of the gauge fields was modified
to be gauge invariant in such a way that it tends to the original mass
term under the Landau gauge. It should be point out with emphasis that
since the mass term of the gauge fields make the theory obey the Lorentz
condition one can regard the theory as a gauge invariant one and correctly
quantize it with the help of such a gauge invariant mass term (see also the
reasoning in setion 2 of the present paper ). As for the renormalisability,
no proof was presented in Ref. [7]. At present the subject can be stated as
follows: Can one prove the renormalisability under the original expression
of the mass term of the gauge fields with a correct quantisation method ?
It will be proven in this paper that the renormalisability of the theory can
be deduced from the renormalisability of the SU(n) Yang-Mills theory. 
\par
     We will use two kinds of path integral of the generating functional for
the Green functions.  One of them consists of the sources associate to all
the variables including the Lagrange multipliers $\lambda_a$. Another one is
the generating functional for the Green functions in the so-called $\xi$
gauge, which does not involve $\lambda_a$. It will be shown that the mass
term of the gauge fields cause no extra complexity to the Slavnov-Taylor
identity of the generating functional $\Gamma$ for the regular vertex
functions and does not change the equations satisfied by the divergent part
of $\Gamma$. Consequently, we will
be able to determine the general form of the counterterms order by order
based on the renormalisability of the Yang--Mills theory and prove that the
mass term of the gauge fields is hurmless to the renormalisability of the
theory. In this way we will also reveal that the renormalisability of the
SU(n) theory with massive gauge bosons is ensured by the renormalisability
of the Yang--Mills theory. The scattering matrix will be discussed in a
separate paper [11].
\par
   The method of quantization will be explained in section $2$. Section $3$ 
and section $4$ are devoted to prove the renormalisability of the theory.
Concluding remarks will be given in the final section.
\par
\vspace{5mm}
\setcounter{equation}{0}
\def\theequation{2.\arabic{equation}}
\begin{center}
{\bf II}.\ \ Quantization and BRST Invariance
\end{center} \par \ \par
     With $A_{a \mu}$, $M$ standing for the SU(n) gauge fields and their
mass parameter the Lagrangian including the mass term ${\cal L}_{AM}$ of
the gauge fields has the form
\begin{eqnarray}
{\cal L} = {\cal L}^{(N)} + {\cal L}_{AM} \,,
\end{eqnarray} 
where
$$
 {\cal L}_{AM} = \frac{1}{2} M^2 A_{a \mu}A_{a}^{\mu} \,,
$$
${\cal L}^{(N)} $ is the Lagrangian of a usual SU(n) gauge theory, namely
\begin{eqnarray*}
&&{\cal L}^{(N)} = -\frac{1}{4} F_{a \mu \nu} F_a^{\mu \nu} 
             + {\cal L}_{\psi} + {\cal L}_{\psi A}  \,, \\
&& F_{a \mu \nu} = \partial_{\mu}A_{a \nu}-\partial_{\nu}A_{a \mu}
                  - g f_{abc} A_{b \mu } A_{c \nu} \,.
\end{eqnarray*}
Under the infinitesimal gauge transformation, one has
\begin{eqnarray*}
\delta \int d^4x {\cal L}(x) &=&\int d^4x  M^2 A_{a \mu}\delta A_{a}^{\mu}
= - \frac{1}{g} M^2 \int d^4x A_{a \mu} \partial^{\mu} \delta \theta_a \\
&=&\, \frac{1}{g} M^2 \int d^4x \big( \partial^{\mu} A_{a \mu}\big)
\delta \theta_a \,,
\end{eqnarray*} 
where $\delta A_a^{\mu}$ stands for the infinitesimal gauge transformation
of the gauge fields
\begin{eqnarray}
 \delta A_a^{\mu} = - \frac{1}{g} \partial^{\mu}\delta \theta_a(x)
       - f_{abc}\delta \theta_b(x) A_c^{\mu}(x) \,.
\end{eqnarray}
     Since the classical equations of motion make the action invariant under
an arbitrary infinitesimal transformation of the field functions, they 
certainly make the mass term of the gauge fields invariant under an arbitrary
infinitesimal gauge transformation. This means that when $M$ is not equal
to zero, the classical equations of motion leads to the following Lorentz
condition 
\begin{eqnarray}
 \partial^{\mu} A_{a \mu} = 0 \,.            
\end{eqnarray}
It should be noticed with emphasis that the Lorentz condition makes the
mass term invariant with respect to the infinitesimal gauge trasformation.
Consequently, the combination of the action and the Lorentz condition is
invariant with respect to the infinitesimal gauge trasformation that
satisfies the following equations
\begin{eqnarray} 
 \delta \big( \partial^{\mu} A_{a \mu} \big) = 0 \,. 
\end{eqnarray} 
\par
    Since such a residual invariance is not broken by the mass term of
the gauge fields it is natural to imagine that the  ghost action should be
the same as that of the SU(n) Yang--Mills theory in the Lorentz gauge (see
for example, Ref. [12]). However, this was often disregarded in the
literature. For intance, in the discussion in Ref. [13], concerning the
massive gauge fields theory without matter fields, the original form of the
generating functional for the Green functions was taken to be
$$ 
 \int {\cal D}[{\cal A}] 
{\rm exp} \{{\rm i} \big[I+ J_a^{\mu}(x)A_{a\mu}(x)\big]\} \,,
$$      
where $J_a^{\mu}(x)A_{a\mu}(x)$ is the source term and $I$ is the action
defined by ${\cal L}(x)$.
In this way, the Lorentz condition (2.3) was ignored. The same drawback
was included in Ref. [14].
\par
    Taking the Lorentz condition into account one should write the path
integral of the Green functions inolving only the original fields as
\begin{eqnarray} 
 \frac{1}{N_0} \int {\cal D}[{\cal A},\overline{\psi},\psi] 
{\Delta}[{\cal A}]
\prod_{a',x'} \delta \left(\partial^{\sigma}A_{a' \sigma}(x') \right)  
A_{a \mu} (x)A_{b \nu}(y)A_{c \rho}(z) \cdots {\rm exp} \{{\rm i} I \} \,,
\end{eqnarray} 
where
$$ 
 N_0 =  \int {\cal D}[{\cal A},\overline{\psi},\psi] \Delta[{\cal A}]
\prod_{a',x'} \delta \left(\partial^{\lambda}A_{a' \lambda}(x') \right) 
 {\rm exp} \{{\rm i} I \} \,.
$$
The problem is to determnined the weight factor $\Delta[{\cal A}]$ and can
be solved by modifying the mass term ${\cal L}_{AM}$ according to the
method of Ref. [7]. In fact, only the field functions which satisfy the
Lorentz condition can play roles in the integral (2.5) and the value of the
Lagrangian can be changed for the field functions which do not satisfy this
condition. In view of the fact that the Lorentz condition makes the mass
term invariant with respect to the infinitesimal gauge trasformation, we
now imagine to replace ${\cal L}_{AM}$ with a gauge invariant mass term
$\widetilde{{\cal L}}_{AM}$ which is equal to ${\cal L}_{AM}$ when the
Lorentz condition is satisfied. Thus, analogous to the case in the
Fadeev--Popov method [1-3,12,13,15], $\Delta[{\cal A}]$ should be gauge
invariant and make the following equation valid for an arbitrary gauge
invariant quantity  ${\cal O}({\cal A,\,\overline{\psi},\,\psi })$
$$
 \int {\cal D}[{\cal A},\,\overline{\psi},\,\psi] 
 \Delta[{\cal A}]
 \prod_{a',x'} \delta \left(\partial^{\lambda}A_{a' \lambda}(x') \right)
 {\cal O}({\cal A},\overline{\psi},\psi)
{\rm exp} \{{\rm i} \widetilde{I} \} 
 \propto  \int {\cal D}[{\cal A},\,\overline{\psi},\,\psi]
{\cal O}({\cal A},\overline{\psi},\psi)
{\rm exp} \{{\rm i} \widetilde{I} \}   \,,
$$
where $\widetilde{I}$ is a gauge invariant action obtained by replacing 
${\cal L}_{AM}$ with $\widetilde{{\cal L}}_{AM}$. This means that 
 $\Delta[{\cal A}]$ can be determined according to the Fadeev--Popov
equation in the usual Yang--Mills theory and is proportional to 
$det[\partial\cdot D]$, where $D$ denots the covariant derivative
in the adjoint representation. Therefore, the ghost Lagrangian or action has
the same form as that of the SU(n) Yang-Mills theory in the Landau gauge.
Namely
\begin{eqnarray}
 {\cal L}^{(C)}(x)
      = \big(- \partial_{\mu}\overline{C}_a(x)\big) D^{\mu}_{ab}C_b(x)  \,, 
  \ \ \ \ \ \
  I^{(C)} = \int d^4x {\cal L}^{(C)}(x)\,,
\end{eqnarray}
where $C_a(x)$ and $\overline{C}_a(x)$ are the F--P ghost fields and
\begin{eqnarray}
 D^{\mu}_{ab}(x) =  \delta_{ab} \partial^{\mu}
                     + g f_{abc} A_c^{\mu}(x)  \,. 
\end{eqnarray}
\par
    As usual one can further generalized the theory by regarding as new
variables the Lagrange multipliers $\lambda_a(x)$ associated with the
Lorentz condition. Thus the total effective Lagrangian and action are
\begin{eqnarray}
&& {\cal L}_{eff}(x) = {\cal L} + {\cal L}^{(C)}(x)
               + \lambda_a(x)\partial^{\mu}A_{a \mu}(x) \,, \\
&& I_{eff} = \int d^4x {\cal L}_{eff}(x) \,. 
\end{eqnarray}
Correspondingly, the path integral of the generating functional for the
Green functions is
\begin{eqnarray}
{\cal Z}[\overline{\eta},\eta,\overline{\chi},\chi, J,j]
= \frac{1}{N_{\lambda}}
 \int {\cal D}[{\cal A},\overline{\psi},\psi,\overline{C},C,\lambda]
 {\rm exp} \Big\{ {\rm i} \big( I_{eff} + I_s \big) \Big\}\,, 
\end{eqnarray}
where $N_{\lambda}$ is a constant to make ${\cal Z}[0,0,0,0,0,0]$ equal
to $1$, $I_s$ is the source term. They are defined by
\begin{eqnarray*}
&& N_{\lambda}
 = \int {\cal D}[{\cal A},\overline{\psi},\psi,\overline{C},C,\lambda]
 {\rm exp} \Big\{ {\rm i} I_{eff}  \Big\} \,, \\
&& I_s = \int d^4x \big[ J_a^{\mu}(x) A_{a \mu}(x) + j_a(x) \lambda_a(x)
    + \overline{\chi}_a(x)C_a(x) \\
&&\ \ \ \ \ \ \ \ \ \ \ \ \ \ \ 
    + \overline{C}_a(x) \chi_a(x)
    + \overline{\eta}_a(x)\psi_a(x) + \overline{\psi}_a(x)\eta_a(x) \big] \,,
\end{eqnarray*}
where
$J_a^{\mu}(x)$, $j_a(x)$, $\overline{\chi}_a(x)$, $\chi_a(x)$ and
$\overline{\eta}$, $\eta$ are the sources associate to various fields.
\par
    We now check the BRST invariance of the effective action $I_{eff}$
defined by (2.8) and (2.9). With the gauge fields, the matter fields and
the ghost fields transforming as usual, one has
\begin{eqnarray*}
&& \delta_B A_a^{\mu} = \delta \zeta D^{\mu}_{ab} C_b(x) \,, \\ 
&& \delta_B {\overline C}_a(x) = - \delta \zeta \lambda_a(x)\,, \\
&& \delta_B C_a(x) = \delta \zeta \frac{g}{2} f_{abc}C_b(x)C_c(x) \,, \\
&& \delta_B I_{eff} 
    =\, \int d^4x \Big \{ \big(
       \delta_B \lambda_a(x) -\delta \zeta M^2 C_a(x)\big)
       \partial^{\mu}A_{a \mu} \Big \} \,,
\end{eqnarray*}
where $\delta \zeta $ is an infinitesimal fermionic parameter independent
of $x$. Obviously, the effective action is invariant when the transformation
of $\lambda_a(x)$  are defined as
$$
 \delta_B \lambda_a(x) =  \delta \zeta M^2 C_a(x) \,.
$$
It is also clear that the transformation is no longer nilpotent.  
\par
    We are also interested in the $\xi$ gauge Green functions that are
defined by replacing the  $\delta-$functions in the numerator and
denominator of (2.5) with the gauge-fixing term
$$  - \frac{1}{2 \xi} (\partial^{\mu}A_{a \mu})^2 \,, $$
where $\xi$ is a pararmter. The total effective Lagrangian and action
including the gauge-fixing term and the ghost term  become (in the same
notations as used above)
\begin{eqnarray}
&& {\cal L}_{eff}(x) = {\cal L} + {\cal L}^{(C)}(x)
              - \frac{1}{2 \xi} (\partial^{\mu}A_{a \mu})^2 \,, \\
&& I_{eff} = \int d^4x {\cal L}_{eff}(x)  \,. 
\end{eqnarray}
Therefore the generating functional for such Green functions is
\begin{eqnarray}  
{\cal Z}[\overline{\eta},\eta,\overline{\chi},\chi, J]
 = \frac{1}{N_{\xi}}\int
{\cal D}[{\cal A},\overline{\psi},\psi, \overline{C}, C]
A_{a \mu} (x)A_{b \nu}(y)A_{b \rho}(z) \cdots
{\rm exp}\Big\{ {\rm i} I_{eff} \Big\} \,,
\end{eqnarray} 
where $N_{\xi}$ is a constant to make ${\cal Z}[0,0,0,0,0]$ equal to $1$,
$I_s$ is the source term 
$$ 
 I_s = \int d^4x \big[ J_a^{\mu}(x) A_{a \mu}(x)  
    + \overline{\chi}_a(x)C_a(x) + \overline{C}_a(x) \chi_a(x) 
    + \overline{\eta}_a(x)\psi_a(x) + \overline{\psi}_a(x)\eta_a(x) \big] \,.
$$
It shuld be noticed that the Lorentz condition will take no effect in the
generating functional for the $\xi$ guage Green functions unless $\xi$ tends
to zero.
\par
\vspace{5mm}
\setcounter{equation}{0}
\def\theequation{3.\arabic{equation}}
\begin{center}
{\bf III}.\ \ Renormalisability 
\end{center} \par
    Based on the quantization method explained in last section we will prove
that the renormalisability of the SU(n) theory with massive gauge bosons can
be deduced from the renormalisability of the Yang--Mills theory. In this
section we will start with the the Green function generating functional
which also includes soruces associated with $\lambda_a$. The method of
reasoning for the theory in the $\xi$ gauge is similar and will be briefly
described in section 4.
\par
    Assume that $A_{a \mu}(x)$, $C_a(x)$, $\overline{C}_a(x)$ and
$\lambda_a(x)$ stand for the renormalized field founctions,  $g$, $M$ are
renormalized parameters. The matter fields $\psi$, $\overline{\psi}$ do not
affect the discussion in this section and will be omitted. As usual we define
the composite field functions $\Delta A_a^{\mu}(x)$ and $\Delta C_a(x)$ by
\begin{eqnarray}
  \delta_B A_a^{\mu}(x) = \delta \zeta \Delta A_a^{\mu}(x)\,,\ \ \ \ \ \
   \delta_B C_a(x) = \delta \zeta \Delta C_a(x) \,,
\end{eqnarray}
where $\Delta A_a^{\mu}(x)$ is just $D^{\mu}_{ab} C_b(x)$ and
 $\Delta C_a(x)$ is $\frac{1}{2} g f_{abc}C_b(x)C_c(x)$. Introducing new
soruces $K^a_{\mu}(x)$ and $L_a(x)$ and adding a soruce term of these
composite fields into the effective Lagrangian without counterterm, one gets
\begin{eqnarray}
 {\cal L}^{[0]}_{eff}(x)  
    =&-&\frac{1}{4} F_{a \mu \nu}(x) F_a^{\mu \nu}(x) 
       + \frac{1}{2} M^2 A_{a \mu}(x) A_a^{\mu}(x) 
     + \lambda_a(x) \partial^{\nu} A_{a \nu}(x) \nonumber\\
     &+& \big(-\partial_{\mu} \overline{C}_a(x)\big) D^{\mu}_{ab}C_b(x)
     \nonumber\\
     &+& K^a_{\mu}(x) \Delta A_a^{\mu}(x) + L_a(x) \Delta C_a(x) \,.
\end{eqnarray}
The complete effective Lagrangian is the sum of ${\cal L}^{[0]}_{eff}$
and the counterterm ${\cal L}_{count}$:
$$
 {\cal L}_{eff} = {\cal L}^{[0]}_{eff} + {\cal L}_{count} \,.
$$
\par
    In terms of the action $I^{[0]}_{eff}$ formed by the effective
Lagrangian ${\cal L}^{[0]}_{eff}$, we define the generating functional
for Green functions
\begin{eqnarray}
{\cal Z}^{[0]}[J,j,\overline{\chi},\chi, K, L]
= \frac{1}{N} \int {\cal D}[{\cal A},\overline{C},C,\lambda]
 {\rm exp} \Big\{ {\rm i} \big( I^{[0]}_{eff} + I_s \big) \Big\}\,, 
\end{eqnarray}
where N is a constant to make ${\cal Z}^{[0]}[0,0,0,0,0,0]$ equal to $1$,
the source term $I_s$ is given by
$$ 
 I_s = \int d^4x \big[ J_a^{\mu}(x) A_{a \mu}(x) + j_a(x) \lambda_a(x)
       + \overline{\chi}_a(x)C_a(x) + \overline{C}_a(x) \chi_a(x) \big]\,.
$$
Correspondingly, the generating functionals ${\cal W}^{[0]}$, $\Gamma^{[0]}$
for connected Green functions and regular vertex functions are 
\begin{eqnarray}
&&{\cal Z}^{[0]}[J,j,\overline{\chi},\chi, K, L] 
  = {\rm exp} \Big\{ {\rm i}{\cal W}^{[0]}[J,j,\overline{\chi},\chi,K,L]
 \Big\} \,, \\
&& \Gamma^{[0]}[\widetilde{A},\widetilde{\overline{C}},\widetilde{C},
   \widetilde{\lambda},K,L]
  = {\cal W}^{[0]}[J,j,\overline{\chi},\chi, K, L] \nonumber\\
&& \ \ \ \ \ \ \ \ \ \ \ \ \ \ \ \
   - \int d^4x \Big[ J_a^{\mu}(x) \widetilde{A}_{a \mu}(x)
       + j_a(x) \widetilde{\lambda}_a(x) 
       + \overline{\chi}_a(x)\widetilde{C}_a(x)
       + \widetilde{\overline{C}}_a(x) \chi_a(x) \Big]\,,
\end{eqnarray}
where $\widetilde{A}_{a \mu}(x)$, $\widetilde{C}_a(x)$,
 $\widetilde{\overline{C}}_a(x)$ and $\widetilde{\lambda}_a(x)$ are the
so-called classical fields defined
by
\begin{eqnarray}
 \widetilde{A}_{a \mu}(x) = \frac{\delta {\cal W}^{[0]}}
                               {\delta J_a^{\mu}(x) } \,, \ \ \ \ \
 \widetilde{\lambda}_a(x) = \frac{\delta {\cal W}^{[0]}}
                               {\delta j_a(x) } \,,\ \ \ \ \ 
 \widetilde{C}_a(x) = \frac{\delta {\cal W}^{[0]}}
                            {\delta \overline{\chi}_a(x)} \,,\ \ \ \ \ 
 \widetilde{\overline{C}}_a(x) = - \frac{\delta {\cal W}^{[0]}}
                                     {\delta \chi_a(x) }\,.
\end{eqnarray}
Therefore one has
\begin{eqnarray}
 J_a^{\mu}(x) = - \frac{\delta \Gamma^{[0]}}
                      {\delta \widetilde{A}_{a \mu}(x) } \,, \ \ \ \ 
 j_a(x) = - \frac{\delta \Gamma^{[0]}}
                      {\delta \widetilde{\lambda}_a(x) } \,, \ \ \ \ 
 \overline{\chi}_a(x) = \frac{\delta \Gamma^{[0]}}
                      {\delta \widetilde{C}_a(x)} \,, \ \ \ \
 \chi_a(x) = - \frac{\delta \Gamma^{[0]}}
                     {\delta \widetilde{\overline{C}}_a(x) }\,,
\end{eqnarray}
and 
\begin{eqnarray}
 \frac{\delta {\cal W}^{[0]}}{\delta K^a_{\mu}(x) }
     = \frac{\delta \Gamma^{[0]}}{\delta K^a_{\mu}(x) } \,, \ \ \ \ \ \ \
 \frac{\delta {\cal W}^{[0]}}{\delta L_a(x)}
     = \frac{\delta \Gamma^{[0]}}{\delta L_a(x)} \,. 
\end{eqnarray}
\par
In order to find the Slavnov--Taylor identity satisfied by the generating
functional for the regular vertex functions, we change the variables in the
path integral of ${\cal Z}^{[0]}$  as follows
\begin{eqnarray*}
&& A_a^{\mu}(x) \rightarrow
   A_a^{\mu}(x) + \delta \zeta \Delta A_a^{\mu}(x)\,, \\
&& C_a(x) \rightarrow
    C_a(x) + \delta \zeta \Delta C_a(x) \,, \\
&& \overline{C}_a(x) \rightarrow  \overline{C}_a(x) 
  - \delta \zeta \lambda_a(x) \,, \\
&& \lambda_a(x) \rightarrow  \lambda_a(x) \,.
\end{eqnarray*}
The volume element of the path integral does not change and  the changes
of the source term and the mass term of the gauge fields  lead to 
\begin{eqnarray}
&& \int d^4x \Big\{ \frac{\delta \Gamma^{[0]}} {\delta K^a_{\mu}(x)}
 \frac{\delta \Gamma^{[0]}} {\delta \widetilde{A}_a^{\mu}(x)} 
+ \frac{\delta \Gamma^{[0]}} {\delta L_a(x)}
 \frac{\delta \Gamma^{[0]}} {\delta \widetilde{C}_a(x)} \nonumber\\
&& \ \ \ \ \ \ \ \ 
    - \widetilde{\lambda}_a(x)
      \frac{\delta \Gamma^{[0]}} {\delta \widetilde{\overline{C}}_a(x)}
  - M^2 \widetilde{A}_{a \mu}(x)
 \frac{\delta \Gamma^{[0]}} {\delta K^a_{\mu}(x)}
 \Big\} = 0 \,.
\end{eqnarray}
Next, by using the invariance of the path integral of ${\cal Z}^{[0]}$
with respect to the translation of the integration variables
 $\overline{C}_a(x)$ and $\lambda_a(x)$, one can get a set of auxiliary
identities 
\begin{eqnarray}
&& \partial_{\mu}\frac{\delta \Gamma^{[0]}} {\delta K^a_{\mu}(x)}
  - \frac{\delta \Gamma^{[0]}}
         {\delta \widetilde{\overline{C}}_a(x)} = 0  \,,
\\
&& \frac{\delta \Gamma^{[0]}} {\delta \widetilde{\lambda}_a(x)}
  - \partial^{\mu} \widetilde{A}_{a \mu}(x) = 0  \,.
\end{eqnarray}
\par
   In the following we will denote
by $ \Gamma^{[0]}[A,\overline{C},C,\lambda,K,L]$ the functional that is
obtained from
$ \Gamma^{[0]}[\widetilde{A},\widetilde{\overline{C}},
\widetilde{C},\widetilde{\lambda},K,L]$
by replacing the classical field functions with the usual field functions.
Defined $\overline{\Gamma}^{[0]}$ as
\begin{eqnarray}
 \overline{\Gamma}^{[0]} = \Gamma^{[0]} -  \int d^4x \Big\{
  \lambda_a(x) \partial^{\mu}A_{a \mu}(x)  \Big\} 
  - \int d^4x \Big\{ \frac{1}{2} M^2 A_a^{\mu}(x) A_{a \mu}(x)  \Big\} \,.
\end{eqnarray}
Thus (3.9)--(3.11) lead to 
\begin{eqnarray}
&& \int d^4x \Big\{
  \frac{\delta \overline{\Gamma}^{[0]}} {\delta K^a_{\mu}(x)}
\frac{\delta \overline{\Gamma}^{[0]}} {\delta A_a^{\mu}(x)}
+ \frac{\delta \overline{\Gamma}^{[0]}} {\delta L_a(x)} 
\frac{\delta \overline{\Gamma}^{[0]}} {\delta C_a(x)}\Big\} = 0 \,,
\\
&& \partial_{\mu}\frac{\delta \overline{\Gamma}^{[0]}} {\delta K^a_{\mu}(x)}
  - \frac{\delta \overline{\Gamma}^{[0]}}
         {\delta \overline{C}_a(x)} = 0  \,,
\\
&&  \frac{\delta \overline{\Gamma}^{[0]}}
         {\delta \lambda_a(x)} = 0  \,.
\end{eqnarray} 
\par
Assume that the dimensional regularization method is used and the relations
(3.13)--(3.15) are guaranteed. Denote the tree part and one loop part
of $\overline{\Gamma}^{[0]}$ by  $\overline{\Gamma}^{[0]}_0$  and
 $\overline{\Gamma}^{[0]}_1$  respectively, $\overline{\Gamma}^{[0]}_0$
 is thus the modified action  $\overline{I}^{[0]}_{eff}$ without the
$\lambda$ term and the mass term of the gauge fields. From (3.13)--(3.15)
one has
\begin{eqnarray}
&& \overline{\Gamma}^{[0]}_0 * \overline{\Gamma}^{[0]}_0 = 0 \,, \\
&& \partial_{\mu}\frac{\delta \overline{\Gamma}^{[0]}_0} {\delta K^a_{\mu}(x)}
 - \frac{\delta \overline{\Gamma}^{[0]}_0}
        {\delta \overline{C}_a(x)} = 0  \,,
\end{eqnarray}
and
\begin{eqnarray}
&& \overline{\Gamma}^{[0]}_0 * \overline{\Gamma}^{[0]}_1
  + \overline{\Gamma}^{[0]}_1 * \overline{\Gamma}^{[0]}_0
  = \Lambda_{\rm op} \overline{\Gamma}^{[0]}_1 = 0 \,,
\\
&& \partial_{\mu}\frac{\delta \overline{\Gamma}^{[0]}_1} {\delta K^a_{\mu}(x)}
 - \frac{\delta \overline{\Gamma}^{[0]}_1}
        {\delta \overline{C}_a(x)} = 0  \,,  \\
&&  \frac{\delta \overline{\Gamma}^{[0]}_1}
         {\delta \lambda_a(x)} = 0  \,.
\end{eqnarray}
The notations  $A*B$, $\Lambda_{\rm op}$ are defined in the usual way,
namely
\begin{eqnarray}
&& A*B = \int d^4x \Big\{
\frac{\delta A} {\delta K^a_{\mu}(x)}
\frac{\delta B} {\delta A_{a \mu}(x)}
+ \frac{\delta A} {\delta L_a(x)} 
  \frac{\delta B} {\delta C_a(x)} \Big\}  \,, \\
&& \Lambda_{\rm op}= \int d^4x \Big\{
\frac{\delta \overline{\Gamma}_0^{[0]}} {\delta K^a_{\mu}(x)}
\frac{\delta}{\delta A_{a}^{\mu}(x)}
+ \frac{\delta \overline{\Gamma}_0^{[0]}} {\delta A_{a}^{\mu}(x)}
\frac{\delta}{\delta K^a_{\mu}(x)}
+ \frac{\delta \overline{\Gamma}_0^{[0]}} {\delta L_a(x)}
\frac{\delta}{\delta C_a(x)}
+ \frac{\delta \overline{\Gamma}_0^{[0]}} {\delta C_a(x)}
 \frac{\delta} {\delta L_a(x)}
 \Big\}  \,. \ \ \ \ \ \ \  
\end{eqnarray}
The pole part of $\overline{\Gamma}^{[0]}_1$ will be denoted by
$\overline{\Gamma}^{[0]}_{1,div}$. Of course it also satisfies
(3.18)--(3.20), namely
\begin{eqnarray}
&& \Lambda_{\rm op} \overline{\Gamma}^{[0]}_{1,div} = 0 \,, \\
&& \partial_{\mu}\frac{\delta \overline{\Gamma}^{[0]}_{1,div}}
                     {\delta K^a_{\mu}(x)}
 - \frac{\delta \overline{\Gamma}^{[0]}_{1,div}}
        {\delta \overline{C}_a(x)} = 0  \,,  \\
&&  \frac{\delta \overline{\Gamma}^{[0]}_{1,div}}
         {\delta \lambda_a(x)} = 0  \,.
\end{eqnarray}
This is the same equations as that appearing in the Yang--Mills theory.
\par
    If $M=0$ then it is known from the renormalisability of the theory that
$ \overline{\Gamma}^{[0]}_{1,div}$ is a combination of the three terms
\begin{eqnarray*}
&& g \frac{\partial \overline{\Gamma}^{[0]}_0}{\partial g}\,,
\ \ \ \ \ \ \ \ \
\int d^4x \Big\{ A_{a \nu}(x)
   \frac{\delta \overline{\Gamma}^{[0]}_0}{\delta A_{a \nu}(x)}
   + L_a(x) \frac{\delta \overline{\Gamma}^{[0]}_0}{\delta L_a(x)}\Big\}\,,
\\
&& \int d^4x \Big\{
    C_a(x) \frac{\delta \overline{\Gamma}^{[0]}_0} {\delta C_a(x) }
 + \overline{C}_a(x) \frac{\delta \overline{\Gamma}^{[0]}_0}
    {\delta \overline{C}_a(x)}
 + K^a_{\mu}(x) \frac{\delta \overline{\Gamma}^{[0]}_0 }{\delta K^a_{\mu}(x)}
    \Big\} \,.
\end{eqnarray*}
Since each of these satisfies equations (3.23)--(3.25) a new term appearing
when $M\not=0$, if any, should includ $M^2$ as a factor and also satisfy
(3.23)--(3.25). Now the equations can not have such a solution. It follows
that
\begin{eqnarray}
&& \overline{\Gamma}^{[0]}_{1,div}
=  \alpha_1 \Big(
             g \frac{\partial \overline{\Gamma}^{[0]}_0}{\partial g} \Big )
+ \beta_1 \int d^4x \Big\{ A_{a \nu}(x)
   \frac{\delta \overline{\Gamma}^{[0]}_0}{\delta A_{a \nu}(x)}
   + L_a(x) \frac{\delta \overline{\Gamma}^{[0]}_0}{\delta L_a(x)}\Big\}
   \nonumber\\
&&\ \ \ \ \ \ \ \ \ \ \ \
 + \gamma_1 \int d^4x \Big\{
    C_a(x) \frac{\delta \overline{\Gamma}^{[0]}_0} {\delta C_a(x) }
 + \overline{C}_a(x) \frac{\delta \overline{\Gamma}^{[0]}_0}
    {\delta \overline{C}_a(x)}
 + K^a_{\mu}(x) \frac{\delta \overline{\Gamma}^{[0]}_0 }{\delta K^a_{\mu}(x)}
    \Big\} \,,
\end{eqnarray}
where $\alpha_1$, $\beta_1$ and $\gamma_1$  are constants of order
$(\hbar)^1$ and are divergent when the space-time dimension tends to $4$. 
\par
    In order to cancel the one loop divergence the counterterm of order
$\hbar^1$ in the action should be chosen as 
\begin{eqnarray}
 \delta I^{[1]}_{count}[A, C, \overline{C}, K, L, g, M] 
  = - \overline{\Gamma}^{[0]}_{1,div}[A, C, \overline{C}, K, L, g, M] \,.
\end{eqnarray}
Thus the sum of $\overline{\Gamma}^{[0]}_0$ and  $\delta I^{[1]}_{count}$,
to order $\hbar^1$, can be written as
\begin{eqnarray}
&& \overline{I}_{eff}^{[1]}[A, C, \overline{C}, K, L, g] \nonumber\\
 &&\ \ \ \ \ \ \ \ 
  = \overline{\Gamma}^{[0]}_0
   [(Z_3^{[1]})^{1/2}A, (\widetilde{Z}_3^{[1]})^{1/2}C,
   (\widetilde{Z}_3^{[1]})^{1/2}\overline{C}, (\widetilde{Z}_3^{[1]})^{1/2}K,
    (Z_3^{[1]})^{1/2}L, Z_g^{[1]} g]  \,,
\end{eqnarray}
where
\begin{eqnarray*}
&& (Z_3^{[1]})^{1/2} = 1 - \beta_1  \,,  \\
&& (\widetilde{Z}_3^{[1]})^{1/2} = 1 - \gamma_1 \,, \\
&& Z_g^{[1]} = 1 - \alpha_1 \,.
\end{eqnarray*}
Next by adding the $\lambda$ term and the mass term of the gauge fields
into $\overline{I}_{eff}^{[1]}$ and forming
\begin{eqnarray} 
 I_{eff}^{[1]}[A, C, \overline{C},\lambda,K, L, g, M] 
 &=& \overline{I}_{eff}^{[1]}[A, C, \overline{C}, K, L, g] \nonumber\\
   && + \int d^4x
   \Big\{ \lambda_a(x) \partial^{\mu} A_{a \mu}(x) \Big\}
   \nonumber\\
   && + \int d^4x\Big\{ \frac{1}{2} M^2 A_{a \mu}(x) A_a^{\mu}(x) \Big\}\,, 
\end{eqnarray}
one has
\begin{eqnarray} 
 I_{eff}^{[1]}[A, C,\overline{C},\lambda,K, L, g, M] 
  = I_{eff}^{[0]}
                 [A^{[0]}, C^{[0]}, \overline{C}^{[0]},\lambda^{[0]},
                 K^{[0]}, L^{[0]}, g^{[0]}, M^{[0]}] \,,
\end{eqnarray}
where $A^{[0]}, C^{[0]}, \overline{C}^{[0]}, \cdots $, to order $\hbar^1$,
stand for the bare quantities and are defined by 
\begin{eqnarray}
&& A^{[0]}_{a \mu} = (Z_3^{[1]})^{1/2} A_{a \mu},\ \ \ \
  C^{[0]}_a = (\widetilde{Z}_3^{[1]})^{1/2}C_a, \ \ \ \
  \overline{C}^{[0]}_a = (\widetilde{Z}_3^{[1]})^{1/2} \overline{C}_a \,,
\\
&& K^{[0]a}_{\mu} = (\widetilde{Z}_3^{[1]})^{1/2} K^a_{\mu},\ \ \ \ \
  L^{[0]}_a = (Z_3^{[1]})^{1/2} L_a  \,,
\\
&&  g^{[0]} = Z_g^{[1]} g,\ \ \ \  M^{[0]} = (Z_3^{[1]})^{-1/2} M ,\ \ \ \
   \lambda_a^{[0]} = (Z_3^{[1]})^{-1/2} \lambda_a \,.
\end{eqnarray}
Obviously, if the action
 $I_{eff}^{[1]}[A, C, \overline{C},\lambda, K, L, g, M]$ is used to replace
  $I_{eff}^{[0]}[A, C,\overline{C},\lambda, K, L, g, M]$ in (3.2)
and define the generating functiomal  $ {\Gamma}^{[1]}$ as well as
$$
 \overline{\Gamma}^{[1]}
  = \Gamma^{[1]}
    - \int d^4x \Big\{
      \lambda_a(x) \partial^{\mu} A_{a \mu}(x) \Big\} 
    -  \int d^4x \Big\{ \frac{1}{2} A_a^{\mu}(x) A_{a \mu}(x)  \Big\} \,,
$$
then one has
\begin{eqnarray}
\overline{\Gamma}^{[1]}[A,\overline{C},C,K,L]
= \overline{\Gamma}^{[0]}[A^{[0]},\overline{C}^{[0]},C^{[0]},K^{[0]},L^{[0]}]
\,.
\end{eqnarray}
We then expand the right hand side of this equation into the form
$$
 \overline{\Gamma}^{[0]}_0[A^{[0]},
\overline{C}^{[0]},C^{[0]},K^{[0]},L^{[0]}]
+ \overline{\Gamma}^{[0]}_1[A^{[0]},
\overline{C}^{[0]},C^{[0]},K^{[0]},L^{[0]}]
+ \cdots \,.
$$
In the first term the divergences of order $\hbar^1$ are due to
 $ \delta I^{[1]}_{count}$. In the second term the divergences of this order
do not contain the contribution of $ \delta I^{[1]}_{count}$ and are
therefore due to the action of order $\hbar^0$. It follows that, to order
$\hbar^1$, $\overline{\Gamma}^{[1]}$ is finite. Moreover from (3.13)--(3.15)
and (3.34) one gets 
\begin{eqnarray}
&& \int d^4x \Big\{
\frac{\delta \overline{\Gamma}^{[1]}} {\delta (K^{[0]})^a_{\mu}(x)}
 \frac{\delta \overline{\Gamma}^{[1]}}{\delta A^{[0]}_{a \mu}(x)} 
 + \frac{\delta \overline{\Gamma}^{[1]}} {\delta L^{[0]}_a(x)} 
\frac{\delta \overline{\Gamma}^{[1]}} {\delta C^{[0]}_a(x)}  \Big\} = 0 \,,
\\
&& \partial_{\mu}\frac{\delta \overline{\Gamma}^{[1]}}
                        {\delta (K^{[0]})^a_{\mu}(x)}
  - \frac{\delta \overline{\Gamma}^{[1]}}
         {\delta \overline{C}_a^{[0]}(x)} = 0  \,, \\
&& \frac{\delta \overline{\Gamma}^{[1]}}
         {\delta \lambda_a^{[0]}(x)} = 0  \,.
\end{eqnarray}
With the help of (3.31)--(3.33), these equations can be written as
\begin{eqnarray}
&& \int d^4x \Big\{
\frac{\delta \overline{\Gamma}^{[1]}} {\delta K^a_{\mu}(x)}
 \frac{\delta \overline{\Gamma}^{[1]}}{\delta A_{a \mu}(x)} 
 + \frac{\delta \overline{\Gamma}^{[1]}} {\delta L_a(x)} 
\frac{\delta \overline{\Gamma}^{[1]}} {\delta C_a(x)} 
\Big\} = 0 \,,  \\
&& \partial_{\mu}\frac{\delta \overline{\Gamma}^{[1]}}
                        {\delta K^a_{\mu}(x)}
  - \frac{\delta \overline{\Gamma}^{[1]}}
         {\delta \overline{C}_a(x)} = 0  \,, \\
&& \frac{\delta \overline{\Gamma}^{[1]}}
              {\delta \lambda_a(x)} = 0  \,.
\end{eqnarray}
\par
    We now know well how to prove the renormalisability of the theory by
using the Slavnov--Taylor identities and the inductive method. Let us
assume that up to $n$ loop the theory has been proved to be renormalisable
by introducing the counterterm
$$
 I^{[n]}_{count} = \sum_{l=1}^{n} \delta I^{[l]}_{count}\,, 
$$
where $\delta I^{[l]}_{count}$ is the counterterm of order $\hbar^l$
and has the form of (3.26),(3.27). This also means that
$\overline{\Gamma}^{[n]}$ determined by the action 
$$ 
I^{[n]}_{eff} = I^{[0]}_{eff} + I^{[n]}_{count} 
$$
satisfies the Slavnov--Taylor identities and is finite to order $\hbar^n$.
We have to proved that by using a counterterm of order $\hbar^{n+1}$
which also has the form of (3.26),(3.27), $\overline{\Gamma}^{[n+1]}$
determined by the action 
$$ 
I^{[n+1]}_{eff} = I^{[n]}_{eff} + \delta I^{[n+1]}_{count} 
$$
can be make satisfy the Slavnov--Taylor identities and finite to
order $\hbar^{n+1}$.
\par
    Denote by $\overline{\Gamma}^{[n]}_k$ the part of order $\hbar^{k}$
in $\overline{\Gamma}^{[n]}$. For $k\leq n$, $\overline{\Gamma}^{[n]}_k$ is
equal to $\overline{\Gamma}^{[k]}_k$, because it can not contain the
contribution of a counterterm of order $\hbar^{k+1}$ or higher.  Thus
on expanding $\overline{\Gamma}^{[n]}$ to order $\hbar^{n+1}$ one has
$$
 \overline{\Gamma}^{[n]}
 = \sum_{k=0}^{n} \overline{\Gamma}^{[k]}_k + \overline{\Gamma}^{[n]}_{n+1}
   + \cdots \,.
$$
Using this and extracting the terms of order $\hbar^{(n+1)}$
in the Slavnov--Taylor identities of $\overline{\Gamma}^{[n]}$,
we find
\begin{eqnarray}
&& \overline{\Gamma}^{[0]}_0 * \overline{\Gamma}^{[n]}_{n+1}
  + \overline{\Gamma}^{[n]}_{n+1} * \overline{\Gamma}^{[0]}_0 = 0 \,,
\\
&& \partial_{\mu}\frac{\delta \overline{\Gamma}^{[n]}_{n+1}}
                                          {\delta K^a_{\mu}(x)}
 - \frac{\delta \overline{\Gamma}^{[n]}_{n+1}}
                      {\delta \overline{C}_a(x)} = 0  \,, \\
&& \frac{\delta \overline{\Gamma}^{[n]}_{n+1}}
                      {\delta \lambda_a(x)} = 0  \,.
\end{eqnarray}
Let $\overline{\Gamma}^{[n]}_{n+1,div}$ stand for the pole part of
$\overline{\Gamma}^{[n]}_{n+1}$. By repeating the steps going from (3.23) 
to (3.26), one can arrive at
\begin{eqnarray}
&& \overline{\Gamma}^{[n]}_{n+1,div}
 = \alpha_{n+1}
      \Big( g \frac{\partial \overline{\Gamma}^{[0]}_0}{\partial g}\Big)
  + \beta_{n+1} \int d^4x \Big\{ A_{a \nu}(x)
   \frac{\delta \overline{\Gamma}^{[0]}_0}{\delta A_{a \nu}(x)}
  + L_a(x) \frac{\delta \overline{\Gamma}^{[0]}_0}{\delta L_a(x)}\Big\}
  \nonumber \\
&& \ \ \ \ \ \ \ \
   + \gamma_{n+1} \int d^4x \Big\{
    C_a(x) \frac{\delta \overline{\Gamma}^{[0]}_0} {\delta C_a(x) }
 + \overline{C}_a(x) \frac{\delta \overline{\Gamma}^{[0]}_0}
    {\delta \overline{C}_a(x)}
 + K^a_{\mu}(x) \frac{\delta \overline{\Gamma}^{[0]}_0 }{\delta K^a_{\mu}(x)}
    \Big\} \,,
\end{eqnarray}
where $\alpha_{n+1}$, $\beta_{n+1}$ and $\gamma_{n+1}$ are
constants of order $(\hbar)^{n+1}$. Therefore, in order to cancel the $n+1$
loop divergence the counterterm of order $\hbar^{n+1}$ should be chosen as 
\begin{eqnarray}
 \delta I^{[n+1]}_{count} [A, C, \overline{C}, K, L, g, M] 
= - \overline{\Gamma}^{[n]}_{n+1,div} [A, C, \overline{C}, K, L, g, M] \,.
\end{eqnarray}
After including this counterterm and the gauge fixing term as well as the
mass term of the gauge feilds $I_{eff}^{[n+1]}$, to order $\hbar^{n+1}$,
can be expressed as
\begin{eqnarray}
 I_{eff}^{[n+1]}[A, C, \overline{C},\lambda, K, L, g, M] 
  = I_{eff}^{[0]}
                 [A^{[0]}, C^{[0]}, \overline{C}^{[0]},\lambda^{[0]},
                 K^{[0]}, L^{[0]}, g^{[0]}, M^{[0]}] \,,
\end{eqnarray}
where $A^{[0]}, C^{[0]}, \overline{C}^{[0]}, \cdots $, to order $\hbar^{n+1}$,
stand for the bare quantities
\begin{eqnarray}
&& A^{[0]}_{a \mu} = (Z_3^{[n+1]})^{1/2} A_{a \mu},\ \ \ \
  C^{[0]}_a = (\widetilde{Z}_3^{[n+1]})^{1/2}C_a, \ \ \ \
  \overline{C}^{[0]}_a = (\widetilde{Z}_3^{[n+1]})^{1/2} \overline{C}_a \,,
\\
&& K^{[0]a}_{\mu} = (\widetilde{Z}_3^{[n+1]})^{1/2} K^a_{\mu},\ \ \ \ \
  L^{[0]}_a = (Z_3^{[n+1]})^{1/2} L_a  \,, \\
&&  g^{[0]} = Z_g^{[n+1]} g,\ \ \ \
   M^{[0]} = (Z_3^{[n+1]})^{-1/2} M ,\ \ \ \
   \lambda_a^{[0]} = (Z_3^{[n+1]})^{-1/2} \lambda_a  \,,
\end{eqnarray}
with
\begin{eqnarray*}
&& (Z_3^{[n+1]})^{1/2}
    = (Z_3^{[n]})^{1/2} - \beta_{n+1} \,,  \\
&& (\widetilde{Z}_3^{[n+1]})^{1/2}
    = (\widetilde{Z}_3^{[n]})^{1/2} - \gamma_{n+1} \,, \\
&& Z_g^{[n+1]} = Z_g^{[n]} - \alpha_{n+1} \,.
\end{eqnarray*}
Therefore, the generating functional $ \overline{\Gamma}^{[n+1]}$ for
proper functions determined by the action $I_{eff}^{[n+1]}$ can
be found from $ \overline{\Gamma}^{[0]}$. Namely 
\begin{eqnarray}
 \overline{\Gamma}^{[n+1]}[A,\overline{C}, C,K,L]
  = \overline{\Gamma}^{[0]}[A^{[0]},\overline{C}^{[0]},
   C^{[0]},K^{[0]},L^{[0]}] \,.
\end{eqnarray}
With this, one can verify that $\overline{\Gamma}^{[n+1]}$ satisfies
(3.38)--(3.40) and is finite to order $\hbar^{n+1}$. Since the theory can
be renormalized to one loop  the renormalisability has been proved by the
inductive method.
\par
\vspace{5mm}
\setcounter{equation}{0}
\def\theequation{4.\arabic{equation}}
\begin{center}
{\bf IV}.\ \ Renormalisability of the theory in the $\xi$ gauge
\end{center} \par
    Similar to section 3, let $A_{a \mu}(x)$, $C_a(x)$
and $\overline{C}_a(x)$ stand for the renormalized field founctions,
$g$, $M$ be renormalized parameters, and $\xi$ is an auxiliary parameter.
The matter fields are also omitted. Now the effective Lagrangian
without counterterm is
\begin{eqnarray}
 {\cal L}^{[0]}_{{\rm eff}}(x)  
    =&-&\frac{1}{4} F_{a \mu \nu}(x) F_a^{\mu \nu}(x) 
       + \frac{1}{2} M^2 A_{a \mu}(x) A_a^{\mu}(x) 
       - \frac{1}{2\xi} \Big(\partial^{\nu} A_{a \nu}(x)\Big)^2 \nonumber \\
     &+& \big(-\partial_{\mu} \overline{C}_a(x)\big) D^{\mu}_{ab}C_b(x)
     \nonumber \\
     &+& K^a_{\mu}(x) \Delta A_a^{\mu}(x) + L_a(x) \Delta C_a(x) \,.
\end{eqnarray}
The generating functional for Green functions is
\begin{eqnarray}
{\cal Z}^{[0]}[J,\overline{\chi},\chi, K, L]
= \frac{1}{N} \int {\cal D}[{\cal A},\overline{C},C]
 {\rm exp} \Big\{ {\rm i} \big( I^{[0]}_{{\rm eff}} + I_s \big) \Big\}\,, 
\end{eqnarray}
where N is a constant to make ${\cal Z}^{[0]}[0,0,0,0,0]$ equal to $1$,
the source term $I_s$ is given by
$$ 
 I_s = \int d^4x \big[ J_a^{\mu}(x) A_{a \mu}(x) 
       + \overline{\chi}_a(x)C_a(x) + \overline{C}_a(x) \chi_a(x) \big]\,.
$$
Correspondingly, the generating functionals ${\cal W}^{[0]}$, $\Gamma^{[0]}$
for connected Green functions and regular vertex functions are 
\begin{eqnarray}
&&{\cal Z}^{[0]}[J,\overline{\chi},\chi, K, L] 
  = {\rm exp} \Big\{ {\rm i}{\cal W}^{[0]}[J,\overline{\chi},\chi,K,L]
 \Big\} \,, \\
&& \Gamma^{[0]}[\widetilde{A},\widetilde{\overline{C}},\widetilde{C},K,L]
  = {\cal W}^{[0]}[J,\overline{\chi},\chi, K, L] \nonumber\\ 
&& \ \ \ \ \ \ \ \ \ \ \ \ \ \ \ \ \ \
- \int d^4x \Big[ J_a^{\mu}(x) \widetilde{A}_{a \mu}(x)
       + \overline{\chi}_a(x)\widetilde{C}_a(x)
       + \widetilde{\overline{C}}_a(x) \chi_a(x) \Big]\,,
\end{eqnarray}
where the classical fields are defined by
\begin{eqnarray}
 \widetilde{A}_{a \mu}(x) = \frac{\delta {\cal W}^{[0]}}
                          {\delta J_a^{\mu}(x) }\,, \ \ \ \ \ \ \
 \widetilde{C}_a(x) = \frac{\delta {\cal W}^{[0]}}
                          {\delta \overline{\chi}_a(x)} \,,\ \ \ \ \ \ \
 \widetilde{\overline{C}}_a(x) = - \frac{\delta {\cal W}^{[0]}}
                                         {\delta \chi_a(x) }\,.
\end{eqnarray}
One therefore has
\begin{eqnarray}
 J_a^{\mu}(x) = - \frac{\delta \Gamma^{[0]}}
                 {\delta \widetilde{A}_{a \mu}(x) } \,, \ \ \ \ \ \ \
 \overline{\chi}_a(x) = \frac{\delta \Gamma^{[0]}}
                 {\delta \widetilde{C}_a(x)} \,, \ \ \ \ \ \ \
 \chi_a(x) = - \frac{\delta \Gamma^{[0]}}
                 {\delta \widetilde{\overline{C}}_a(x) }\,,
\end{eqnarray}
and 
\begin{eqnarray}
\frac{\delta {\cal W}^{[0]}}{\delta K^a_{\mu}(x) }
        = \frac{\delta \Gamma^{[0]}}{\delta K^a_{\mu}(x) } \,, \ \ \ \ \ \ \
 \frac{\delta {\cal W}^{[0]}}{\delta L_a(x)}
        = \frac{\delta \Gamma^{[0]}}{\delta L_a(x)} \,.
\end{eqnarray}
\par
In order to find the Slavnov--Taylor identity satisfied by the generating
functional for the regular vertex functions, we change the variables in the
path integral of ${\cal Z}^{[0]}$  as follows
\begin{eqnarray*}
&& A_a^{\mu}(x) \rightarrow
   A_a^{\mu}(x) + \delta \zeta \Delta A_a^{\mu}(x)\,, \\
&& C_a(x) \rightarrow
    C_a(x) + \delta \zeta \Delta C_a(x) \,, \\
&& \overline{C}_a(x) \rightarrow  \overline{C}_a(x) 
  + \delta \zeta \frac{1}{\xi} \partial_{\mu} A_a^{\mu}(x) \,.
\end{eqnarray*}
The volume element of the path integral does not change and  the changes
of the source term and the mass term of the gauge fields  lead to 
\begin{eqnarray}
&& \int d^4x \Big\{ 
\frac{\delta \Gamma^{[0]}} {\delta K^a_{\mu}(x)}
 \frac{\delta \Gamma^{[0]}} {\delta \widetilde{A}_a^{\mu}(x)} 
+ \frac{\delta \Gamma^{[0]}} {\delta L_a(x)}
 \frac{\delta \Gamma^{[0]}} {\delta \widetilde{C}_a(x)} \nonumber\\
&& \ \ \ \ \ \ \ \ \ \ 
 + \frac{1}{\xi}\big(\partial_{\mu} \widetilde{A}_a^{\mu}(x)\big)
    \frac{\delta \Gamma^{[0]}} {\delta \widetilde{\overline{C}}_a(x)}
  - M^2 \widetilde{A}_{a \mu}(x)
 \frac{\delta \Gamma^{[0]}} {\delta K^a_{\mu}(x)}
 \Big\} = 0 \,.
\end{eqnarray}
Next, by using the invariance of the path integral of ${\cal Z}^{[0]}$
with respect to the translation of the integration variables
 $\overline{C}_a(x)$, one can get a set of auxiliary identities 
\begin{eqnarray}
 \partial_{\mu}\frac{\delta \Gamma^{[0]}} {\delta K^a_{\mu}(x)}
  - \frac{\delta \Gamma^{[0]}}
         {\delta \widetilde{\overline{C}}_a(x)} = 0  \,.
\end{eqnarray}
\par
    Let $\Gamma^{[0]}[A,\overline{C},C,K,L]$ be the functional
that is obtained from
$ \Gamma^{[0]}[\widetilde{A},\widetilde{\overline{C}},\widetilde{C},K,L]$
by replacing the classical field functions with the usual field functions.
Defined $\overline{\Gamma}^{[0]}$ as
\begin{eqnarray}
 \overline{\Gamma}^{[0]} = \Gamma^{[0]} + \int d^4x \Big\{
   \frac{1}{2\xi} \Big(\partial^{\nu} A_{a \nu}(x)\Big)^2 \Big\}
   - \int d^4x \Big\{ \frac{1}{2} M^2 A_{a \mu}(x) A_a^{\mu}(x)
    \Big\} \,.
\end{eqnarray}
Thus from (4.8) and (4.9) one has
\begin{eqnarray}
&& \int d^4x \Big\{
  \frac{\delta \overline{\Gamma}^{[0]}} {\delta K^a_{\mu}(x)}
\frac{\delta \overline{\Gamma}^{[0]}} {\delta A_a^{\mu}(x)}
+ \frac{\delta \overline{\Gamma}^{[0]}} {\delta L_a(x)} 
\frac{\delta \overline{\Gamma}^{[0]}} {\delta C_a(x)}\Big\} = 0 \,, \\
&& \partial_{\mu}\frac{\delta \overline{\Gamma}^{[0]}} {\delta K^a_{\mu}(x)}
  - \frac{\delta \overline{\Gamma}^{[0]}}
         {\delta \overline{C}_a(x)} = 0  \,.
\end{eqnarray}
\par
It is now obvious that the method used in last section can be followed
to prove the renormalisability of the theory in the $\xi$ gauge.
\par
\vspace{5mm}
\begin{center}
{\bf V}.\ \ Concluding Remarks 
\end{center} \par
     We have expounded that the quantization under the Lorentz condition
caused by the mass term of the gauge fields leads to a ghost action which is
the same as that of the usual SU(n) Yang--Mills theory in the Landau gauge.
    Furthermore, we have clarified that the mass term of the gauge fields
cause no extra complexity to the Slavnov-Taylor identity of the generating
functional for the regular vertex functions. In particular, the equations
satisfied by the divergent part of this generating functional are independent
of $M$. Consequently, we have been able to determine the general form of the
counterterms order by order based on the renormalisability of the Yang--Mills
theory and prove that the mass term of the gauge fields is hurmless to
the renormalisability of the theory. In this way we have also revealed that
the renormalisability of the SU(n) theory with the mass term of the gauge
fields is ensured by that of the Yang--Mills theory theory.
\par
\vspace{4mm}
\begin{center}
\bf{ACKNOWLEDGMENTS}
\end{center} \par
    We are grateful to Professor Yang Li-ming for 
helpful discussions. This work was supported in part by National Natural 
Science Foundation of China and  supported in part by Doctoral Programm
Foundation of the Institution of Higher Education of China.
\par
\ \par
\begin{center}
{\large \bf Refernces}
\end{center}
\par  \noindent
[1]\ L.D. Faddeev and V.N. Popov, Phys. Lett. {\bf B25}, 29(1967).  
\par  \noindent
[2]\ B.S. De Witt, Phys. Rev. {\bf 162},1195,1239(1967).
\par  \noindent
[3]\ L.D. Faddev and A.A.Slavnov, Gauge Field: Introduction to Quatum Theory,
\par\ \ \ \  The Benjamin Cummings Publishing Company, 1980.
\par  \noindent
[4]\ G.Curci and R.Ferrari, Nuovo Cim. {\bf 32}, 151(1976).
\par  \noindent
[5]\ I.Ojima, Z. Phys. {\bf C13},173(1982).
\par  \noindent
[6]\ A.Blasi and N.Maggiore, het-th/9511068; Mod. Phys. Lett.{\bf A11},
1665(1996).
\par  \noindent
[7]\ R.Delbourgo and G.Thompson, Phys. Rev. Lett. {\bf 57}, 2610(1986).
\par  \noindent
[8]\ M.Carena and C.Wagner, Phys. Rev. {\bf D37}, 560(1988).
\par  \noindent
[9]\ A.Burnel, Phys. Rev. {\bf D33}, 2981(1986);{\bf D33}, 2985(1986);.
\par  \noindent
[10]\ T.Fukuda, M.Monoa, M.Takeda and K.Yokoyama, \par\ \ \ \ 
Prog. Theor. Phys. {\bf 66},1827(1981);{\bf 67},1206(1982);{\bf 70},284(1983).
\par  \noindent
[11]\ Ze-sen Yang, Xianhui Li, Zhining Zhou and Yushu Zhong, hep-th/9912034.
\par  \noindent 
[12]\ G.H.Lee and J.H.Yee, Phys. Rev. {\bf D46}, 865(1992).
\par  \noindent
[13]\ C.Itzykson and F-B.Zuber, Quantum Field Theory, McGraw-Hill, 
      New York, 1980.
\par  \noindent
[14]\ C.Grosse-Knetter, Phys. Rev. {\bf D48}, 2854(1993).
\par  \noindent
[15]\ Ze-sen Yang, Advanced Quantum Machanics, Peking University Press,
     2-ed.  Beijing, 1995. 
\par
\end{document}